\documentclass[a4paper,twocolumn,11pt,accepted=2022-05-22]{quantumarticle}
\pdfoutput=1
\usepackage[utf8]{inputenc}
\usepackage[english]{babel}
\usepackage[T1]{fontenc}
\usepackage{amsmath}
\usepackage{hyperref}
\usepackage[numbers,sort&compress]{natbib}

\begin{document}

\title{A general quantum algorithm for open quantum dynamics demonstrated with the Fenna-Matthews-Olson complex}

\author{Zixuan Hu}
\affiliation{Department of Chemistry, Department of Physics, and Purdue Quantum Science and Engineering Institute, Purdue University, West Lafayette, IN 47907, USA}
\orcid{0000-0002-0752-3811}
\author{Kade Head-Marsden}
\affiliation{John A. Paulson School of Engineering and Applied Sciences, Harvard University, Cambridge, MA 02138, USA}
\orcid{0000-0002-4122-4081}
\author{David A. Mazziotti}
\affiliation{Department of Chemistry and The James Franck Institute, The University of Chicago, Chicago, IL 60637 USA}
\orcid{0000-0002-9938-3886}
\author{Prineha Narang}
\affiliation{John A. Paulson School of Engineering and Applied Sciences, Harvard University, Cambridge, MA 02138, USA}
\orcid{0000-0003-3956-4594}
\author{Sabre Kais}
\email{kais@purdue.edu}
\affiliation{Department of Chemistry, Department of Physics, and Purdue Quantum Science and Engineering Institute, Purdue University, West Lafayette, IN 47907, USA}
\orcid{0000-0003-0574-5346}

\maketitle

\begin{abstract}
  Using quantum algorithms to simulate complex physical processes and correlations in quantum matter has been a major direction of quantum computing research, towards the promise of a quantum advantage over classical approaches. In this work we develop a  generalized quantum algorithm to simulate any dynamical process represented by either the operator sum representation or the Lindblad master equation. We then demonstrate the quantum algorithm by simulating the dynamics of the Fenna-Matthews-Olson (FMO) complex on the IBM QASM quantum simulator. This work represents a first demonstration of a quantum algorithm for open quantum dynamics with a moderately sophisticated dynamical process involving a realistic biological structure. We discuss the complexity of the quantum algorithm relative to the classical method for the same purpose, presenting a decisive query complexity advantage of the quantum approach based on the unique property of quantum measurement. 
\end{abstract}

\section{Introduction}

Simulating physical processes with quantum algorithms has been a major focus of quantum computing research~\cite{Kais2014, Cao2019, Bauer2020, HeadMarsdenFlick2020, Huang2020, Childs2018}. Open quantum dynamics studies the time evolution of a quantum system interacting with an environment~\cite{Breuer2002}. The complexity of the environment often makes exact treatment impractical, and various approximation approaches have been developed to treat the environment as an averaging effect on the system of interest. So far relatively few studies have been done to develop quantum algorithms for open quantum dynamics~\cite{Wang2011, Wang2013, Wei2016, DiCandia2015, Sweke2015, Garcia-Perez2020, Childs2016, Sweke2015, Kliesch2011} despite its importance in modeling realistic physical systems. One main challenge is the time evolution of open quantum systems is non-unitary, while quantum algorithms are realized by unitary quantum gates. To tackle this problem an early study~\cite{Wang2011} added an auxiliary environment to the system thus making the total evolution unitary. Mimicking the Markovian process, that procedure required a reset of the auxiliary environment at each time step, which could be difficult to achieve if the system and the environment are entangled or the evolution time is long. Other studies~\cite{Wang2013, Wei2016} have proposed novel quantum algorithms to simulate open quantum dynamics, yet the demonstrations are limited to basic 2-level systems. In this work we propose a general quantum algorithm that can be demonstrated with a much more complex and realistic physical model. In a previous study we have developed a quantum algorithm for evolving the dynamics in the operator sum representation~\cite{Hu2020}. The algorithm uses the Sz.-Nagy dilation approach to convert each Kraus operator into its corresponding unitary dilation matrix, which is then implemented on a quantum circuit. The quantum algorithm was successfully applied to the amplitude damping quantum channel and implemented on the IBM Q quantum simulator and quantum computers~\cite{Hu2020}. The quantum algorithm was designed with generality in mind and indeed has been adapted through use of an ensemble of Lindbladian trajectories method~\cite{HeadMarsden2019PRA, HeadMarsden2019JCP} to simulate non-Markovian dynamics as demonstrated on the Jaynes-Cummings model~\cite{HeadMarsden2020}. However, there are two issues that must be solved before our dilation-based quantum algorithm can be applied to more complex dynamical processes. Firstly, the Kraus operators in the amplitude damping model have an explicit dependence on time, which allows each time step to be simulated independently. This explicit dependence however cannot be easily obtained for more complex dynamical models. Secondly, a large class of dynamical models are described by the master equation formulation. To apply our quantum algorithm to such models a connection between the master equation and the operator sum representation must be established. Although there have been studies of converting a widely used type of master equation – the Lindblad equation – into the operator sum representation~\cite{Nakazato2006, Andersson2007}, those methods solve the master equation into a time-explicit matrix form and therefore require integration of the superoperators describing the dynamics. Integration of the superoperators is hard for classical algorithms and currently not solvable with any quantum algorithm. In the following we intentionally avoid integrating the superoperators, but instead treat the master equation with the Euler method -- this is an essential simplification that allows practical application of the quantum algorithm to more sophisticated dynamical models.

In this work we present a generalized quantum algorithm that can simulate any open quantum dynamics represented by either the operator sum representation or the Lindblad master equation. In this algorithm each Kraus operator can be directly related to a physical process without the integration of superoperators. In addition, the Kraus operators no longer require an explicit dependence on time and the system at a given time is evolved from the initial state by iteratively applying the Kraus operators obtained from the physical processes. To demonstrate the generality of the generalized algorithm, we use it to simulate a moderately sophisticated dynamical process for a realistic biological structure: the Fenna-Matthews-Olson (FMO) complex. The Fenna-Mathews-Olson complex is a trimeric-pigment protein complex found in green sulfur bacteria~\cite{Blankenship2014}. In the photosynthetic light-harvesting process, it is responsible for transferring excitonic energy from the antennae complexes to the reaction center with nearly 100\% efficiency~\cite{Andrews1999, Sension2007,BarrosoFlores2017}. While this is a very well-studied complex~\cite{Ishizaki2009a, Ishizaki2009b, Zhu2012, Thyrhaug2018, IrgenGioro2019, Oh2019, Suzuki2020, Kim2020}, a more in-depth understanding of this transport process can provide valuable insight for the dynamics of other light harvesting complexes or for the design of artificial photosynthetic systems~\cite{Yeh2012, Hu2018}. 

While some earlier studies exist~\cite{Gupta2020, Mahdian2020a, Mahdian2020b}, the application of our generalized quantum algorithm to the FMO dynamics as demonstrated on the IBM QASM simulator~\cite{Qiskit} is so far as we know the first successful quantum simulation of a moderately sophisticated dynamical process involving a realistic biological structure. The simulation showcases the generality of the method by mapping the Lindblad master equation of the FMO into the operator sum representation. The same as previously discussed in Ref.~\cite{Hu2020}, in terms of the gate count required to execute the evolution, the generalized quantum algorithm has comparable computational complexity as classical methods. However, here we emphasize that both the previous and the generalized quantum algorithms can have a decisive query complexity advantage over any classical method, when evaluating an observable over the density matrix. Under specific conditions the query complexity advantage can even translate to the total complexity advantage. This complexity advantage is the result of the fundamentally quantum properties of superposition and projection measurement. 

\section{Results}
\subsection{Generalized Quantum Algorithm}
\label{sec:improved_algo}

The basic mechanisms of the generalized quantum algorithm -- including the Sz.-Nagy unitary dilation procedures and the observable evaluation procedures -- have been introduced in our previous work~\cite{Hu2020} and reviewed briefly below:

We assume the physical composition of the initial density matrix is known and can be expressed by a sum of different pure quantum states weighted by the corresponding probabilities:
\begin{equation}
    \rho = \sum_i p_i \lvert \phi_i \rangle \langle \phi_i \rvert
    \label{eq:1}
\end{equation}
where each $p_i$ is the probability of finding each $\lvert \phi_i \rangle$ in the mixed state of $\rho$. Now if the dynamical model is given by the operator sum representation: 
\begin{equation}
    \rho(t) = \sum_k M_k \rho M_k^{\dagger}
    \label{eq:2}
\end{equation}
we want to simulate the time evolution of $\rho(t)$ given the initial $\rho$ and the Kraus operators $M_k$’s. To achieve this, we can prepare each $\lvert \phi_i \rangle$ as an input state vector $v_i$ in a given basis and then use a quantum circuit to create the quantum state: 
\begin{equation}
    \lvert \phi_{ik}(t) \rangle = M_kv_i \xrightarrow[\mathrm{dilation}]{\mathrm{unitary}}U_{M_k}(v_i^T,0,...,0)^T
    \label{eq:3}
\end{equation}
where $U_{M_k} = \begin{pmatrix} M_k & D_{M_k^{\dagger}}\\
D_{M_k} & -M_k^{\dagger}
\end{pmatrix}$  is the 1-dilation of  $M_k$~\cite{Levy2014, Hu2020}, ${{D}_{{{M}_{k}}}}=\sqrt{I-M_{k}^{\dagger }{{M}_{k}}}$ is the defect operator of $M_k$.  After $\lvert \phi_{ik}(t) \rangle$ has been obtained for each $M_k$ and each $v_i$, the population of each basis state in the current basis can be obtained by calculating the diagonal vector:
\begin{equation}
    \mathrm{diag}(\rho(t)) = \sum_{ik} p_i  \cdot \mathrm{diag}(\lvert \phi_{ik} \rangle \langle \phi_{ik} \rvert)
    \label{eq:4}
\end{equation}
where $\mathrm{diag}(\lvert \phi_{ik}(t) \rangle \langle \phi_{ik}(t) \rvert)$ can be efficiently obtained by applying projection measurements on the first half subspace of $U_{M_k}(v_i^T,0,...,0)^T$.

To evaluate the expectation value of an observable $\langle A \rangle = \mathrm{Tr}(A\rho(t))$, we consider the operator $\tilde{A} = \frac{A + \mathbf{I}\lvert\lvert A\rvert\rvert}{2\lvert\lvert A\rvert\rvert}$ with the Cholesky decomposition $\tilde{A} = LL^{\dagger}$~\cite{Menon2013}. We can then evolve:
\begin{equation}
    L^{\dagger}\lvert \phi_{ik}(t) \rangle = L^{\dagger}M_kv_i \xrightarrow[\mathrm{dilation}]{\mathrm{unitary}}U_{L^{\dagger}}U_{M_k}(v_i^T,...,0)^T
    \label{eq:5}
\end{equation}
and obtain $\langle \tilde{A} \rangle$ by:
\begin{equation}
    \langle \tilde{A} \rangle = \mathrm{Tr}(\tilde{A}\rho(t)) =  \sum_{i,k}\mathrm{Tr}(p_i\cdot L^{\dagger} \lvert \phi_{ik}(t) \rangle \langle \phi_{ik}(t) \rvert L)
    \label{eq:6}
\end{equation}
where the trace of $L^{\dagger} \lvert \phi_{ik}(t) \rangle \langle \phi_{ik}(t) \rvert L$ can be obtained by projection measurements into the first $N$-dimensional space of $U_{L^{\dagger}}U_{M_k}(v_i^T,0,...,0)^T$~\cite{Hu2020}. After this, $\langle A \rangle$ can then be obtained from $\langle \tilde{A} \rangle$ by $\langle A \rangle  = 2\lvert\lvert A \rvert\rvert \langle \tilde{A} \rangle - \lvert\lvert A \rvert\rvert$. This procedure was then applied to the amplitude damping channel: 
\begin{align}
    \label{eq:7}
    \rho(t) &= M_0\rho M_0^{\dagger} + M_1\rho M_1^{\dagger}\\
    \notag
    M_0 &=\begin{pmatrix}
    1 & 0\\
    0 & \sqrt{e^{-\gamma t}}
    \end{pmatrix}\\
    \notag
    M_1 &= \begin{pmatrix}
    0 & \sqrt{1-e^{-\gamma t}}\\
    0 & 0
    \end{pmatrix}
\end{align}
where $M_0$ and $M_1$ both have an explicit dependence on time $t$. This concludes the review of the previous algorithm, for more details please see Ref.~\cite{Hu2020}.

Now to introduce the new features that allow the generalized algorithm to simulate more complex dynamics, we first note that the Kraus operators with explicit time dependence as reviewed in Equation~\eqref{eq:7} are not generally available for more complex dynamics. This is especially true when the dynamics is described by a master equation, where the superoperators have to be integrated first to obtain the Kraus operators with explicit time dependence. To avoid the hard problem of integrating superoperators, a naive and incorrect idea is to directly write the Kraus operators of an arbitrary dynamics in the same basic forms of those used in Equation~\eqref{eq:7} simulating the amplitude damping channel. However as shown in the Supplementary Information Section S1 for the finite-temperature amplitude damping channel, Kraus operators with naive dependence on time lead to incorrect dynamics for even such a simple deviation from the amplitude damping channel. Indeed, the original formulation of the operator sum representation~\cite{Nielsen2011} can be essentially understood as a single physical event that starts with an initial density matrix $\rho(0)$ and ends with a final density matrix $\rho(1)$: $\rho(1) = \mathcal{E}_1[\rho(0)] = \sum_k M_{1k}\rho(0)M_{1k}^{\dagger}$, where $\mathcal{E}_1$ is the quantum operation described by the collection of $M_{1k}$’s. Now if we use $\rho(1)$ as the initial state and apply another quantum operation, we have $\rho(2) = \mathcal{E}_2[\rho(1)] = \sum_k M_{2k}\rho(1)M_{2k}^{\dagger}$. This process can be repeated iteratively such that: 
\begin{equation}
    \rho(s) = \mathcal{E}_s[\rho(s-1)] = \sum_k M_{sk}\rho(s-1)M_{sk}^{\dagger}
    \label{eq:8}
\end{equation}
where each $\rho(s)$ with $s=1,2,...,S$ can be understood as a discrete time step that samples the dynamics until we reach the final time step. Now an outstanding question is how to determine the Kraus operators $M_{sk}$ for each quantum operation $\mathcal{E}_s$. The answer is if the time evolution involves the same physical processes throughout the total time interval being simulated, then we can use the physical processes to determine the collection of $M_k$’s and use the same collection for all the iterations. For example, in the amplitude damping channel, the physical process is a single transition from the excited state to the ground state, and the corresponding Kraus matrix is $M_1 = \begin{pmatrix} 0 & \sqrt{p}\\ 0 & 0 \end{pmatrix}$, where $p$ is the probability of the transition happening over a single time step. The condition $\sum_k M_k^{\dagger}M_k = \mathbf{I}$ thus leads to $M_0 = \begin{pmatrix} 1 & 0\\ 0 & \sqrt{1-p}\end{pmatrix}$ . Applying the iterative process in Equation~\eqref{eq:8} to an initial $\rho(0)$ with the same $M_0$ and $M_1$ for each iteration we have the excited state population at the final time step:
\begin{align}
    \label{eq:9}
    \rho_{11}(S) &= \rho_{11}(0)(1-p)^S\\
    \notag
    &= \rho_{11}(0) (1-\gamma\delta t)^{\frac{t}{\delta t}} \xrightarrow{\delta t \rightarrow 0} \rho_{11}(0)e^{-\gamma t}
\end{align}
where we assume $p$ is given by a rate constant $\gamma$ times a small time interval $\delta t$ and $t=S\delta t$. Equation~\eqref{eq:9} demonstrates that when the Kraus operators obtained from the physical processes are applied iteratively to the initial density matrix, we obtain the expected exponential decay for the excited population under the condition that the time interval $\delta t$ between each time step is small compared to the total time $t$. It is easy to verify that the correct dynamics is also obtained for all other entries of the density matrix. In the Supplementary Information Section S1 the same procedure also produces the correct dynamics for the finite temperature amplitude damping. The relation between the transition probability and the rate constant in Equation~\eqref{eq:9} also points to a natural relation between the operator sum representation and the Lindblad master equation. Indeed the Lindblad master equation with the form of 
\begin{equation}
    \frac{d\rho(t)}{dt} = \sum_{k > 0} \gamma_k\big(L_k\rho(t)L_k^{\dagger} - \frac{1}{2}\{L_k^{\dagger}L_k, \rho(t)\}\big)
    \label{eq:10}
\end{equation}
describes each physical process with the Lindblad operator $L_k$ and the rate constant $\gamma_k$  (the $\{ \cdot , \cdot \}$ is an anticommutator). Now if we associate each $L_k$ with an $M_k$, with the details of the physical process described by the form of the operator $L_k$, and the probability of the physical process given by $p_k = \gamma_k\delta t$, then~\cite{Breuer2002}: 
\begin{equation}
    M_k = \sqrt{\gamma_k \delta t}L_k
    \label{eq:11}
\end{equation}
Now setting $M_0 = \mathbf{I} - \frac{1}{2}\delta t \sum_{k>0} \gamma_k L_k^{\dagger} L_k$, using $\rho(t)$ as the initial state and incrementing time by $\delta t$ we get:
\begin{widetext}
\begin{align}
    \label{eq:12}
    \rho(t+\delta t) &= M_0 \rho(t) M_0^{\dagger} + \sum_{k>0}M_k\rho(t)M_k^{\dagger}\\ 
    \notag
    &= \rho(t) - \frac{1}{2}\delta t \sum_{k>0} \gamma_k \{L_k^{\dagger}L_k,\rho(t)\} + \mathcal{O}(\delta t^2) + \delta t\sum_{k>0} \gamma_kL_k\rho(t)L_k^{\dagger}
\end{align}
\end{widetext}
Equation~\eqref{eq:12} converges to Equation~\eqref{eq:10} when $\delta t \rightarrow 0$. At the same time: 
\begin{widetext}
\begin{align}
    \label{eq:13}
    \sum_{k} M_k^{\dagger}M_k &= M_0^{\dagger}M_0 + \sum_{k>0} M_k^{\dagger}M_k\\
    \notag
    &= \mathbf{I} - \delta t\sum_{k>0}\gamma_k L_k^{\dagger}L_k + \mathcal{O}(\delta t^2) + \delta t \sum_{k>0} \gamma_k L_k^{\dagger}L_k\\
    \notag
    &= \mathbf{I} + \mathcal{O}(\delta t^2)
\end{align}
\end{widetext}
Therefore the condition of $\sum_{k} M_k^{\dagger}M_k = \mathbf{I}$ is approximately satisfied for the operator sum representation. Equations~\eqref{eq:10} through \eqref{eq:13} show that the Lindblad master equation and the operator sum representation are two different descriptions of the same dynamics: when the physical processes are represented by the Lindblad operators $L_k$’s, the Lindblad master equation defines a differential equation on the density operator $\rho(t)$; when the physical processes are represented by the Kraus operators $M_k$’s, the operator sum representation essentially evolves the differential equation with the Euler method. In actual simulation of the dynamics, Equation~\eqref{eq:11} is especially important as it allows us to easily determine the Kraus operators $M_k$’s with $k>0$ for use in the iterative process in Equation~\eqref{eq:8}. In Equation~\eqref{eq:13} the condition $\sum_{k} M_k^{\dagger}M_k = \mathbf{I}$ is not exactly satisfied, which may cause problems in our specific quantum algorithm. However this problem is easy to solve as we can enforce the condition by defining:
\begin{equation}
    M_0 = \sqrt{\mathbf{I} - \sum_{k>0} M_k^{\dagger}M_k}
    \label{eq:14}
\end{equation}
Now with all the Kraus operators defined, we can simulate the dynamics by the iterative process in Equation~\eqref{eq:8} with the same Kraus operators for each iteration. The first three time steps will look like: 
\begin{widetext}
\begin{align}
\label{eq:15}
    \rho(1) = \sum_k M_k\rho(0)M_k^{\dagger} &\\
    \notag
    \rho(2) = \sum_k M_k\rho(1)M_k^{\dagger} &= \sum_j\sum_k M_jM_k\rho(0)M_k^{\dagger}M_j^{\dagger}\\
    \notag
    \rho(3) = \sum_k M_k\rho(2)M_k^{\dagger} &= \sum_i\sum_j\sum_k M_iM_jM_k\rho(0)M_k^{\dagger}M_j^{\dagger}M_i^{\dagger}
\end{align}
\end{widetext}
For the first time step we can apply the basic unitary dilation procedure described in Equations~\eqref{eq:1} to~\eqref{eq:6} or in Ref.~\cite{Hu2020} to implement each $M_k$ term. For the second and later time steps we need to use higher dilations~\cite{Hu2020, Levy2014} to implement each term, e.g. a 2-dilation for each $M_jM_k$ and a 3-dilation for each $M_iM_jM_k$. In Equation~\eqref{eq:15} if the total number of Kraus operators is $K$ – meaning there are $(K-1)$ physical processes – for the first time step we need to implement $K$ $M_k$ terms, for the second step $K^2$ $M_jM_k$ terms, for the third step $K^3$ $M_iM_jM_k$ terms, and in general we need to implement $K^s$ terms for the $s^{th}$ time step. This exponential scaling of terms with the number of time steps apparently limits the total number of time steps that can be simulated. In actual simulation however, there are two ways the number of terms can be significantly reduced such that the total number of time steps simulated can be greatly increased (See the Methods section for details). As shown in Equation~\eqref{eq:12} the time interval $\delta t$ between time steps must be kept small for the Euler method to work, therefore increasing the total number of time steps is equivalent to increasing the total time for which the dynamics can be evolved. In the following our quantum simulation covers sufficient total time to observe the FMO dynamics. If after all the simplifications the total time is still too short for observing some other physical phenomena, quantum tomography may be applied to the final $S^{th}$ $\rho(S)$ which can be then used as the new initial state. Quantum tomography however will likely take a lot of computational resource and will be beyond the scope of the study.

\subsection{Simulation of the Fenna-Matthews-Olson Dynamics}
\label{sec:fmo}

\begin{figure}[h]
\centering
 \includegraphics[width=0.25\textwidth]{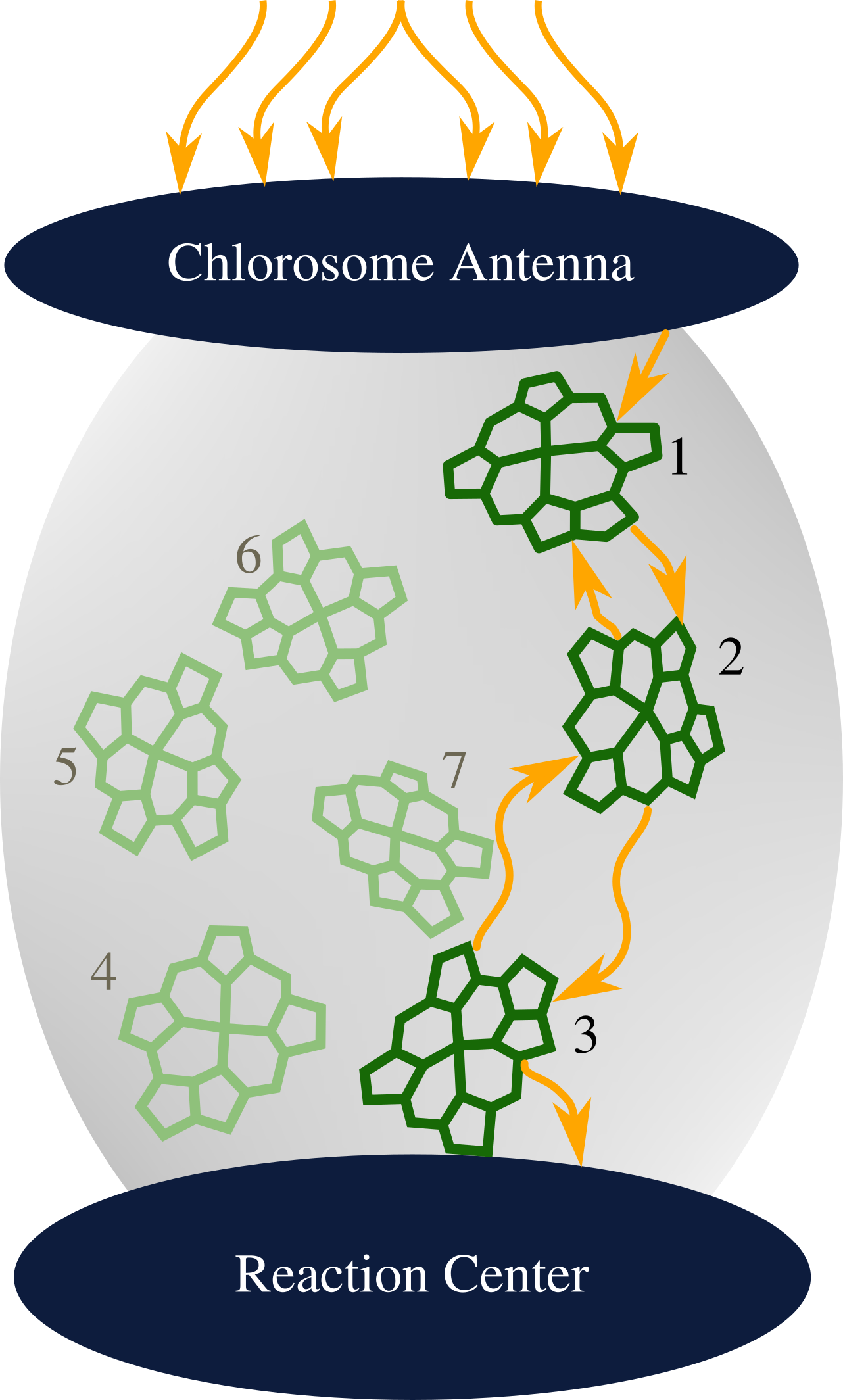}\\
 \caption{One monomer of the Fenna-Matthews-Olsen complex where the seven-chromophore system acts as the quantum wire connecting the chlorosome antenna to the reaction center. The three-chromophore functional subsystem is presented in green, with yellow arrows showing the interactions between the chromophores.}
 \label{fig:fmo}
\end{figure}

In the photosynthetic light-harvesting process, the FMO is a trimer complex which acts as a quantum wire connecting the light-harvesting antennae to the reaction center~\cite{Engel2007, Scholes2017}. Each monomer consists of seven bacteriochlorophyll chromophores, where an initial excitation can occur on either chromophore 1 or 6  and is transported to chromophore 3, which is closely coupled to the reaction center~\cite{Adolphs2006, Skochdopole2011}. This transport is largely driven by environmental interactions including those with neighboring excitations and the protective protein scaffold~\cite{Valleau2017}. Within each monomer of this trimer, there are multiple efficient quantum pathways for the exciton to be transferred to the reaction center. This quantum redundancy has led to a study of the different functional subsystems of the FMO’s chromophores, demonstrating that many subsets exist with similar efficiencies to that of the entire monomer~\cite{Skochdopole2011}. In this work, we consider the three-chromophore subsystem which consists of chromophores 1-3 and has been shown to faithfully reproduce the exciton dynamics on the entire seven-chromophore monomer~\cite{Skochdopole2011}. A single FMO monomer is shown in Figure~\ref{fig:fmo}, where the functional subsystem is depicted in green with the interactions depicted with yellow arrows. 
Considering the redundancy of the chromophore subunits in the FMO dynamics, we can focus on three chromophore subunits supporting three local states of excitation. Adding a ground state and a sink state, the Hamiltonian of the relevant part of the FMO is:
\begin{equation}
    H = \sum_{i=0}^4 \omega_i \sigma_i^+ \sigma_i^- + \sum_{j \neq i} J_{ij}(\sigma_i^+\sigma_j^- + \sigma_j^+\sigma_i^-)
    \label{eq:16}
\end{equation}
where the state $\lvert i \rangle$ with energy $\omega_i$ is created by the Pauli raising operator $\sigma_i^+$ and annihilated by the Pauli lowering operator $\sigma_i^-$, $J_{ij}$ is the coupling strength between $\lvert i \rangle$ and $\lvert j \rangle$. The FMO dynamics can be described by the Lindblad master equation: 
\begin{equation}
    \frac{d\rho(t)}{dt} = -i[H,\rho(t)] + \sum_{k>0} \big( L_k\rho(t)L_k^{\dagger} - \frac{1}{2}\{L_k^{\dagger}L_k, \rho(t)\}\big)
    \label{eq:17}
\end{equation}
where the $[\cdot,\cdot ]$ is a commutator and the $\{\cdot, \cdot \}$ is an anticommutator, and the seven $L_k$’s represent seven physical processes (the rate $\gamma_k$ has been folded into each $L_k$ as defined next). 

\begin{figure}[h]
\centering
 \includegraphics[width=1.1\columnwidth]{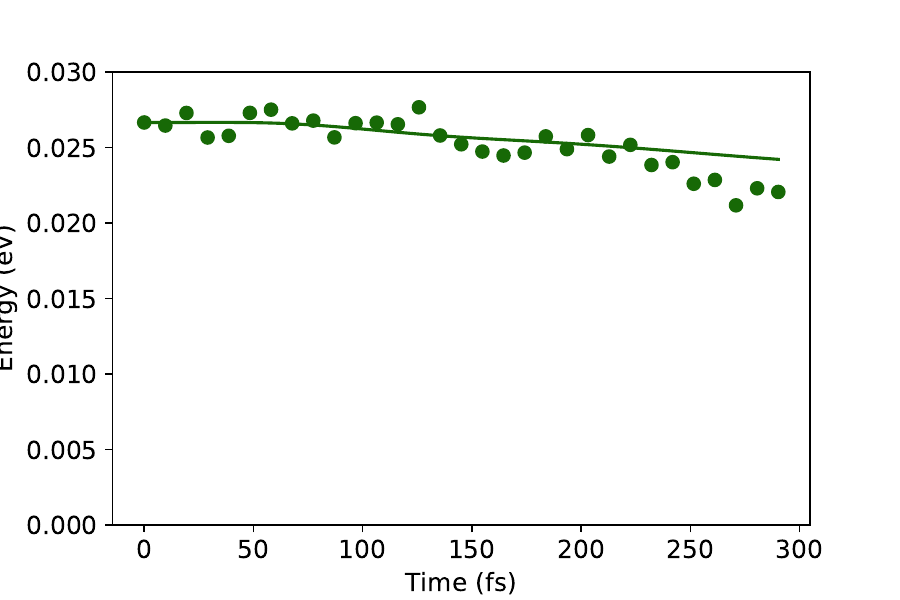}\\
 \caption{The dynamical evolution of the energy observable in the FMO complex model. The classical benchmark is shown as a smooth curve, and the quantum simulation results are shown as dots. For the quantum results, five groups of simulations with six data points each are run, creating totally 30 data points evenly spaced within the 290 fs total time. Each data point is the average of 9216 measurement results. An example of the quantum circuits can be found in the Supplementary Section S5.
}
 \label{fig:energy}
\end{figure}

$L_1$ through $L_3$ are the dephasing operators $L_{deph}(i) = \sqrt{\alpha} \lvert i \rangle \langle i \rvert$ with $i=1,...,3$; $L_4$ through $L_6$ are the dissipation operators that describe the transition from $\lvert i \rangle$ to the ground $\lvert 0 \rangle$: $L_{diss}(i) = \sqrt{\beta} \lvert 0 \rangle \langle i \rvert$ with $i=1,...,3$; $L_{sink}$ is the sink operator that describes the transition from $\lvert 3 \rangle$ to the sink $\lvert 4 \rangle$: $L_{sink} = \sqrt{\gamma} \lvert 4 \rangle \langle 3 \rvert $. Compared to Equation~\eqref{eq:10}, Equation~\eqref{eq:17} contains an additional term of $-i[H,\rho(t)]$. This “coherent part” of the dynamics is unitary and thus can be easily realized by multiplying each Kraus operator by an additional unitary matrix.  All the parameters used in the Hamiltonian and the Lindblad master equation are listed in the Supplementary Information Section S2. Now to determine the Kraus operators we use Equations~\eqref{eq:11} and ~\eqref{eq:14}: 
\begin{align}
\label{eq:18}
    M_1 &= \sqrt{\alpha\delta t} \lvert 1 \rangle \langle 1 \rvert &
    M_2 &= \sqrt{\alpha\delta t} \lvert 2 \rangle \langle 2 \rvert \\
    \notag
    M_3 &= \sqrt{\alpha\delta t} \lvert 3 \rangle \langle 3 \rvert &
    M_4 &= \sqrt{\beta\delta t} \lvert 0 \rangle \langle 1 \rvert \\
    \notag
    M_5 &= \sqrt{\beta\delta t} \lvert 0 \rangle \langle 2 \rvert &
    M_6 &= \sqrt{\beta\delta t} \lvert 0 \rangle \langle 3 \rvert\\
    \notag
    M_7 &= \sqrt{\gamma\delta t} \lvert 4 \rangle \langle 3 \rvert &
    M_0 &= \sqrt{\mathbf{I} - \sum_{k>0} M_k^{\dagger}M_k}
\end{align}
where the time interval $\delta t$ is set to be 2000 atomic unit or 48.4 fs. We then apply the procedure in Equation~\eqref{eq:15} for up to the 6$^{th}$ time step with a total simulation time of 12000 atomic unit or 290 fs. In principle, with eight Kraus operators, the total number of terms from Equation~\eqref{eq:15} for the 6$^{th}$ iteration would be a prohibitively large number $8^6 =  262144$. However by considering redundancy of the terms and setting a threshold of the norm of the matrices to be $>0.01$ (reduction of terms discussed in the Methods section), we are able to reduce the number of terms to only 679, which is much more manageable. We then construct the quantum circuits from the unitary dilations of the Kraus matrices(for detailed procedures see Equations~\eqref{eq:1} to~\eqref{eq:6} or Ref.~\cite{Hu2020}) and run the circuits on the IBM QASM simulator~\cite{Qiskit}. An example of the circuits is shown in the Supplementary Section S5. The results are shown in Figure~\ref{fig:dynamics} and Figure~\ref{fig:energy}:
\begin{figure}[h]
\centering
 \includegraphics[width=1.1\columnwidth]{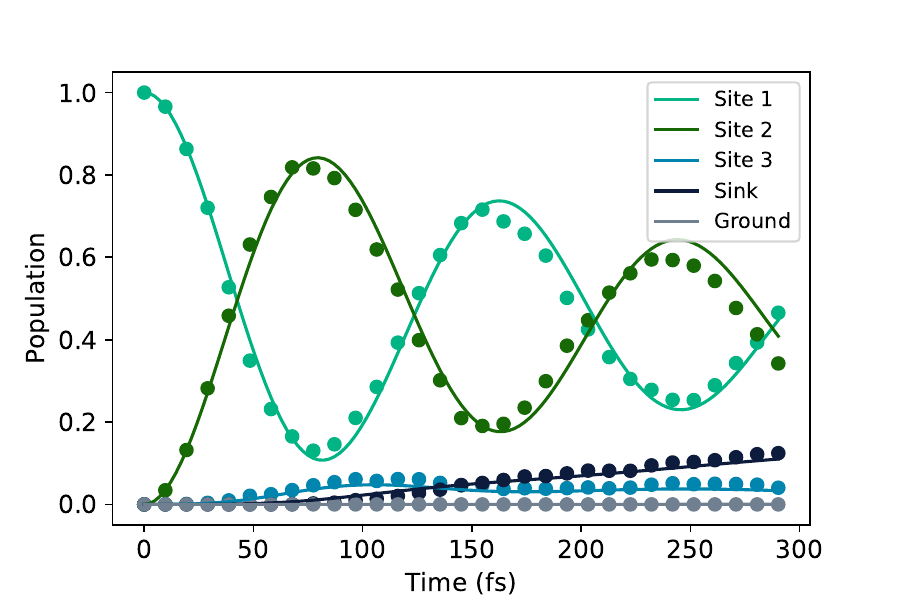}\\
 \caption{The dynamical evolution of the populations of the five states in the FMO complex model. The classical benchmarks are shown as smooth curves, and the quantum simulation results are shown as dots. In the quantum results, for each population, five groups of simulations with six data points each are run, creating totally 30 data points evenly spaced within the 290 fs total time. Each data point is the average of 9216 measurement results. An example of the quantum circuits can be found in the Supplementary Section S5.}
 \label{fig:dynamics}
\end{figure}

In Figure~\ref{fig:dynamics} we see the dynamical evolution of the populations of all the five states in our model. These populations are calculated by the procedure explained around Equation~\eqref{eq:4}, where the diagonal elements of the evolved density matrix are obtained by projection measurements into the computational subspace. We first simulate six data points at 2000 (48.4 fs), 4000 (96.8 fs), 6000 (145 fs), 8000 (194 fs), 10000 (242 fs) and 12000 (290 fs) atomic unit respectively. To obtain more data points such that the oscillations can be smoothly represented, we simulate four additional groups with six data points each (details in the Methods section). The results simulated by our generalized quantum algorithm as implemented on the IBM QASM simulator are shown as dots and they agree well with the classically calculated results shown as curves. The results in Figure~\ref{fig:dynamics} demonstrate the viability of the generalized quantum algorithm. In particular the iterative procedure in Equation~\eqref{eq:8} , the formulation of Kraus operators from Lindblad operators in Equations~\eqref{eq:11} and ~\eqref{eq:14}, and the simplification of terms by norm threshold and redundancy, are working together to produce the correct population dynamics with small errors. In Figure~\ref{fig:dynamics} we can see the excitation beating between chromophores 1 and 2, as it gradually decays into the sink. This process is driven by a combination of environmental noise and entanglement between the chromophores in the functional subsystem.  

In Figure~\ref{fig:energy}, we see the expectation values of the energy observable calculated at different time steps. The evaluation of an observable by projection measurement is a complex process as explained in Equations~\eqref{eq:5},~\eqref{eq:6} and Ref.~\cite{Hu2020}. Again, the quantum simulation results shown as dots agree well with the classical results shown as a curve. Indeed, the way we evaluate the observable in the generalized algorithm is the same as the previous quantum algorithm~\cite{Hu2020}, therefore if the new algorithm can accurately simulate the evolution of the density matrix, then the observable evluation will not introduce new errors not considered before.

\section{Analysis of complexity scaling and error}
\label{sec:comperror}

\subsection{The complexity scaling}
\label{sec:comp}

On the complexity scaling of the quantum algorithm, there are two separate scalings for the evolution of the dynamics: the scaling with the system size $n$ and the scaling with the evolution time step.

For the scaling with the system size $n$, for an arbitrary $n\times n$  Kraus operator ${{\mathbf{M}}_{k}}$ without any special property, the cost to realize its unitary dilation ${{\mathbf{U}}_{{{\mathbf{M}}_{k}}}}$ is $\mathcal{O}\left( {{n}^{2}} \right)$~\cite{Hu2020} . In practice however, an ${{\mathbf{M}}_{k}}$ is often sparse with few non-zero elements. This is particularly true for our current quantum algorithm as we generate the ${{\mathbf{M}}_{k}}$’s from the basic physical processes of transition and dephasing. Consequently the complexity scaling of a quantum circuit implementation for each ${{\mathbf{M}}_{k}}$ can be greatly reduced to $\mathcal{O}\left( {{\log }^{2}}n \right)$ ~\cite{Hu2020}. Now we still need to evolve all $K$ number of ${{\mathbf{M}}_{k}}$’s, so the total complexity scaling is $\mathcal{O}\left( K{{\log }^{2}}n \right)$, where $K$ is determined case-by-case by the dynamical model. Note that the different ${{\mathbf{M}}_{k}}$’s can be evolved in parallel, thus the scaling in $K$ is a “soft” scaling because it does not contribute to either the depth or the width of each individual quantum circuit. 

For the scaling with the evolution time step, without any simplification, Step s requires that ${{K}^{s}}$ matrices be evolved, which is an exponential scaling in the time step. Fortunately, the number of terms can be reduced drastically by the sparsity and redundancy of the ${{\mathbf{M}}_{k}}$ matrices that we discussed in detail under Equation~\eqref{eq:15} and in the Methods section. In general it is difficult to claim the scaling is polynomial or exponential, because the number of matrix terms depends, in a case-by-case matter, on the actual forms of the ${{\mathbf{M}}_{k}}$ matrices and the threshold set for the matrix norm. For our particular FMO model we have managed to simplify the number of terms from the prohibitive ${{8}^{6}}=262144$ to merely 679, which allowed us to implement it on the IBM simulator. Here we remark that the scaling with the evolution time step is the natural result of solving a differential equation with the Euler method. The Euler method is the simplest among the Runge-Kutta methods that are standard tools for the numerical evolution of differential equations. Consequently, the scaling with the evolution time is not an artifact introduced by our method, but an important issue to be investigated if one wishes to adapt classical numerical tools for differential equations to the quantum computing field. With case-specific simplifications to reduce the scaling, our method represents a step in this direction.

So far we have discussed the two complexity scalings for the evolution of the dynamics. In the quantum computing context, the complexity scaling associated with the measurement and retrieval of results – the query complexity – is also important. Here we emphasize that both the previous~\cite{Hu2020, HeadMarsden2020} and the current generalized versions of the quantum algorithm can have a decisive polynomial-versus-exponential query complexity advantage over classical methods when used to evaluate an observable over the density matrix. In addition, in specific situations where the evolution complexity is below a threshold, the total complexity is dominated by the query complexity, and therefore the quantum algorithm can have a decisive total complexity advantage over any classical methods. This quantum advantage along with an example system is discussed in greater detail in the Supplementary Information Section S4.

\subsection{The error analysis}
\label{sec:error}

There are two main sources of errors in the current study. The first source is the error introduced by evolving the master equation with the Euler method and this error is well known to be $\mathcal{O}\left( \delta {{t}^{2}} \right)$. As a future direction, this error could be potentially reduced by introducing high-order methods such as the Runge-Kutta methods. Note that this source of error is not unique to our quantum algorithm but is also present in many classical algorithms used to numerically solve a differential equation. The second source is the measurement error when the IBM QASM simulator simulates a quantum projection measurement. This error is notably only scaling inversely with the number of measurements taken, and not scaling with the system size, as we have discussed in Section S3 of the Supplementary Information. This error is also not unique to our quantum algorithm but always present when one wants to extract physical information from the evolved density matrix. Overall, these two errors did not prevent us from obtaining reasonably good results compared to the classical benchmarks. A third source of error – the gate error – would become important when one tries to run the quantum algorithm on an actual quantum computer. Unfortunately the quantum computers available to us at the writing of the manuscript have non-trivial gate errors such that an experimental demonstration of our quantum algorithm has not been achieved. Nonetheless, the demonstration on the IBM QASM simulator shows that our quantum algorithm is fully programmable (without blind spots or “oracles”) and cost-efficient (with case-specific simplifications), such that it can be implemented on quantum computers with smaller gate error or better error-mitigation capabilities in the near future.

\section{Conclusions and Outlook} 
\label{sec:conc}

In this work we have developed a generalized quantum algorithm for open quantum dynamics to simulate more complex dynamical models. In particular we formulate the Kraus operators from basic physical processes that constitute the dynamics, and then realize the time evolution by an iterative process. We then relate the Lindblad operators to the Kraus operators in a simple manner with the Lindblad master equation being connected to the operator sum representation through the correspondence between a differential equation and its integration by the Euler method. The generalized quantum algorithm works with any dynamical model represented by the operator sum representation or the Lindblad master equation, and thus is much more general than the previous algorithm. We demonstrate the generalized algorithm on the FMO dynamical model using the IBM QASM simulator, which is so far as we know the first successful quantum simulation of a moderately sophisticated dynamical process involving a realistic biological structure. Finally we analyze the query complexity of the algorithm, and constructs an example where the quantum algorithm showcases decisive complexity advantage over any classical method.

\section{Methods} 
\label{sec:meth}

\subsection{Reduction of terms in Equation~\eqref{eq:15}}

To reduce the number of terms in Equation~\eqref{eq:15} and increase the total simulation time, the first way is to notice that the $M_k$’s with $k>0$ typically correspond to basic physical processes such as state transition or dephasing, and thus an $M_k$ typically contains only one or few non-zero elements. For example, a transition from the state $\lvert 1 \rangle$ to $\lvert 0 \rangle$ corresponds to the operator $\lvert 0 \rangle \langle 1 \rvert$ whose matrix $M$ has only one non-zero element at the $M[0,1]$ location regardless of the total dimension of the matrix. Matrices with few non-zero elements multiplying each other will often produce the zero matrix or matrices with negligible norms. Hence by setting a threshold for the norm of the matrix, we can significantly reduce the number of terms to implement as in Equation~\eqref{eq:15} while maintaining a reasonable accuracy. The second way for term reduction is to notice that many of the $K^s$ terms for the $s^{th}$ time step are different by only a constant. This is again due to the fact that the $M_k$’s typically contain few non-zero elements, such that repeated multiplications of certain $M_k$’s tend to form closed groups. Hence by organizing the terms in Equation~\eqref{eq:15} into types within which the matrices only differ by a constant, we can also significantly reduce the number of terms to implement. The actual effects of both ways of term reduction are case specific, and we have shown the numbers in the FMO simulation section.

\subsection{Simplified procedure for implementing products of Kraus operators}

To implement each term in Equation~\eqref{eq:15} such as the $M_iM_jM_k$ for the 3$^{rd}$ time step, we need to use a 3-dilation unitary having four times the dimension of $M_k$, and this increases the gate count of the quantum circuit. Here expecting a large gate count that may exceed the capability of the current noisy intermediate-scaled quantum (NISQ) devices~\cite{Preskill2018quantumcomputingin}, we calculate the product of the matrices such as $M_iM_jM_k$ classically and only implement the product as a single 1-dilation unitary. Note that the sparsity of the $M_k$ matrices ensures the classical calculation of the product is simple, and with more powerful quantum devices in the future we can easily switch back to an all quantum implementation of the matrix products with higher dilations.

\subsection{Procedure to obtain more data points}

To obtain more data points such that the oscillations can be smoothly represented, we simulate four additional groups with six data points each: e.g. the first group starting with 400 atomic unit (9.68 fs) and incrementing by 2000 atomic unit (48.4 fs): 400 (9.68 fs), 2400 (58.1 fs), 4400 (106 fs), 6400 (155 fs), 8400 (203 fs), 10400 (252 fs) atomic unit, the second group starting with 800 atomic unit (19.4 fs) and incrementing by 2000 atomic unit: 800 (19.4 fs), 2800 (67.7 fs), 4800 (116 fs), 6800 (164 fs), 8800 (213 fs), 10800 (261 fs), etc.. In this way we essentially simulate five independent groups, which when put into the same figure, represent 30 data points equally spaced within the 12000 atomic unit (290 fs) total time.

\begin{acknowledgments}
\noindent \textbf{Acknowledgments} 
All authors acknowledge the funding by the U.S. Department of Energy (Office of Basic Energy Sciences) under Award No. DE-SC0019215. All authors acknowledge the use of IBM Quantum services for this work. The views expressed are those of the authors, and do not reflect the official policy or position of IBM or the IBM Quantum team.

\end{acknowledgments}

\bibliographystyle{unsrtnat}
\bibliography{main}




  

\onecolumn\newpage
\appendix

\setcounter{equation}{0}
\setcounter{figure}{0}
\setcounter{table}{0}
\setcounter{page}{1}
\setcounter{section}{0}
\makeatletter
\renewcommand{\theequation}{S\arabic{equation}}
\renewcommand{\thesection}{S\arabic{section}}
\renewcommand{\thefigure}{S\arabic{figure}}
\renewcommand{\bibnumfmt}[1]{[#1]}
\renewcommand{\citenumfont}[1]{#1}

\section*{Supplementary Information}

\section{The finite-temperature amplitude damping channel treated by the operator sum representation}
\label{sec:amp_damp}
In the main text we showed the Kraus matrices used to simulate the amplitude damping channel in Equation~\eqref{eq:7},
\begin{align}
    \label{eq:S1}
    \tag{S1}
    \rho(t) &= M_0\rho M_0^{\dagger} + M_1\rho M_1^{\dagger}\\
    \notag
    M_0 &=\begin{pmatrix}
    1 & 0\\
    0 & \sqrt{e^{-\gamma t}}
    \end{pmatrix}\\
    \notag
    M_1 &= \begin{pmatrix}
    0 & \sqrt{1-e^{-\gamma t}}\\
    0 & 0
    \end{pmatrix}
\end{align}
where $M_0$ and $M_1$ both have an explicit dependence on time $t$. Now the finite-temperature amplitude damping channel includes one more physical process than the amplitude damping channel: the transition from the ground state back to the excited state. If we formulate the dynamics of the channel with a naive dependence on time as in Equation~\ref{eq:S1}, we would have:
\begin{align}
    \tag{S2}
    \label{eq:S2}
    \rho(t) &= M_0\rho M_0^{\dagger} + M_1\rho M_1^{\dagger} + M_2\rho M_2^{\dagger}\\
    \notag
    M_0 &= \begin{pmatrix} \sqrt{e^{-\gamma_2 t}} & 0\\ 0 & \sqrt{e^{-\gamma_1 t}}\end{pmatrix}\\ 
    \notag
    M_1 &= \begin{pmatrix} 0 & \sqrt{1-e^{-\gamma_1 t}}\\ 0 & 0\end{pmatrix}\\ 
    \notag
    M_2 &= \begin{pmatrix} 0 & 0\\ \sqrt{1-e^{-\gamma_2 t}} & 0\end{pmatrix}
\end{align}
where $\gamma_1$ and $\gamma_2$ are the rates of the two transitions. Now the population of the ground state will be:
\begin{align}
    \tag{S3}
    \label{eq:S3}
    \rho_{00}(t) &= (1-e^{-\gamma_1 t})\rho_{11}(0) + e^{-\gamma_2 t}\rho_{00}(0)\\
    \notag
    \rho_{00}(\infty) &= \rho_{11}(0)
    \notag
\end{align}
which is incorrect because the textbook solution of this model is: 
\begin{align}
    \label{eq:S4}
    \tag{S4}
    \rho_{00}(t) &= e^{-(\gamma_1+\gamma_2) t}\rho_{00}(0) + \frac{\gamma_1}{\gamma_1 + \gamma_2}\big(1- e^{-(\gamma_1 + \gamma_2) t}\big)\\
    \notag
    \rho_{00}(\infty) &= \frac{\gamma_1}{\gamma_1 + \gamma_2}
    \notag
\end{align}
Hence we see that Kraus operators formulated with a naive dependence on time lead to incorrect dynamics for even such a simple deviation from the amplitude damping channel. 

Now to treat the dynamics correctly we use the iterative procedure in Equation~\eqref{eq:8} in the main text:
\begin{align}
    \tag{S5}
    \label{eq:S5}
    \rho(s) &= \mathcal{E}_s[\rho(s-1)] = \sum_{k=0}^{2}M_{sk}\rho(s-1)M_{sk}^{\dagger}\\
    \notag
    M_{s0} &= \begin{pmatrix} \sqrt{1-p_2} & 0 \\ 0 & \sqrt{1-p_1} \end{pmatrix}\\
    \notag
    M_{s1} &= \begin{pmatrix} 0 & \sqrt{p_1}\\ 0 & 0 \end{pmatrix}\\
    \notag
    M_{s2} &= \begin{pmatrix} 0 & 0\\ \sqrt{p_2} & 0 \end{pmatrix}
    \notag
\end{align}
where $p_1 = \gamma_1\delta t$ and $p_2 = \gamma_2\delta t$ are the probabilities of the corresponding transition happening over a small time interval $\delta t$. Now the population of the ground state after $S$ iterations is:
\begin{align}
    \tag{S6}
    \label{eq:S6}
    \rho_{00}(S) &= (1-p_1-p_2)^S\rho_{00}(0) + \sum_{i=0}^{S-1}p_1(1-p_1-p_2)^i\\
    \notag
    &= (1-p_1-p_2)^S\rho_{00}(0) + \frac{p_1}{p_1+p_2}[1-(1-p_1-p_2)^S]
\end{align}
Recognizing that $\delta t = \frac{t}{S}$, substituting in  $p_1 = \gamma_1\delta t$ and $p_2 = \gamma_2\delta t$, we have:
\begin{align}
    \tag{S7}
    \label{eq:S7}
    \rho_{00}(t) &= (1-(\gamma_1 + \gamma_2)\frac{t}{S})^S\rho_{00}(0) + \frac{\gamma_1}{\gamma_1+\gamma_2}[1-(1-(\gamma_1+\gamma_2)\frac{t}{S})^S]\\
    \notag
    &\xrightarrow{S\rightarrow \infty} e^{-(\gamma_1+\gamma_2)t}\rho_{00}(0) + \frac{\gamma_1}{\gamma_1 + \gamma_2}[1-e^{-(\gamma_1+\gamma_2)t}]
\end{align}
which is the correct solution as in Equation~\ref{eq:S4}. Therefore we see that the iterative procedure described in Equation~\eqref{eq:8} in the main text will give the correct solution if $\delta t$ is small compared to the total time $t$.

\section{Parameters used for the FMO dynamics simulation in the main text.}

The matrix form of the Hamiltonian of the FMO in the unit of eV:  
\begin{equation}
    \tag{S8}
    \label{eq:S8}
    H_{FMO} = \begin{pmatrix} 0 & 0 & 0 & 0 & 0\\ 0 & 0.0267 & -0.0129 & 0.000632 & 0\\ 0 & -0.0129 & 0.0273 & 0.00404 & 0\\ 0 & 0.000632 &  0.00404 & 0 & 0\\ 0 & 0 & 0 & 0 &0\end{pmatrix}
\end{equation}

\begin{table}[h!]
    \centering
    \begin{tabular}{c|c|c}
    \hline\hline 
    Parameter & Value & Symbol\\
    \hline
    Dephasing rate & $\alpha$ & 3.00x10$^{-3}$fs$^{-1}$\\
    Dissipation rate & $\beta$ & 5.00x10$^{-7}$fs$^{-1}$\\
    Sink & $\gamma$ & 6.28x10$^{-3}$fs$^{-1}$\\
    \hline\hline
    \end{tabular}
    \caption{Rate constants used in the 3-site Fenna-Matthews-Olson complex model.}
    \label{tab:my_label}
\end{table}

\section{Number of projection measurements required to achieve a certain error limit.}
\label{sec:S3}

A quantum state $\lvert \phi \rangle = (c_1,c_2,...,c_n)$ defines a probability distribution of the basis states (e.g. $(1,0,0,..,0)^T$ , $(0,1,0,..,0)^T$ , etc.) with the probability of getting the $i^{th}$ basis state equal to $\lvert c_i \rvert^2$. Applying projection measurements on such a quantum state is essentially sampling this probability distribution. It is a basic result that if we want to deduce the original probability distribution with some finite number of sampling measurements, then the error decreases with increasing number of measurements. In particular, if the error of the sampling can be represented by the standard error of the mean $\sigma_{mean}$, and the original probability distribution defines a standard deviation $\sigma$, then:
\begin{align}
    \tag{S9}
    \label{eq:S9}
    \sigma_{mean} &= \frac{\sigma}{\sqrt{P}}\\
    \notag
    P &= \big(\frac{\sigma}{\sigma_{mean}}\big)^2
    \notag
\end{align}
where $P$ is the number of measurements. We then have the result mentioned in the following Section~\ref{sec:S4} that $P$ does not scale with the dimension $n$ of the quantum state, but only depends on the error $\sigma_{mean}$ we can tolerate. 

\section{Quantum advantage in query complexity}
\label{sec:S4}
Compared to the previous quantum algorithm, the generalized quantum algorithm generalizes to more complex dynamical models by utilizing an iterative process as in Equation~\eqref{eq:8} in the main text and formulating the Kraus operators with physical processes as in Equation~\eqref{eq:11}in the main text. Therefore, at the core of the generalized algorithm is still implementing matrices multiplying vectors by unitary dilations and an execution complexity comparable to classical methods is maintained at $\mathcal{O}(n^2)$ ($n$ being the dimension of the vectors used to represent the initial states). However, here we emphasize that both versions of the quantum algorithm can have a significant query complexity advantage over classical methods, when used to evaluate an observable over the density matrix. As shown in Equations~\eqref{eq:5} and ~\eqref{eq:6} in the main text, after the observable  $A$ has been converted into $\tilde{A} = L^{\dagger}L$, we can evaluate $\langle A \rangle$ by calculating the trace $\mathrm{Tr}\big(p_i\cdot L^{\dagger}\lvert \phi_{ik}(t) \rangle\langle \phi_{ik}(t)\lvert L\big)$ and then summing over $i$ and $k$. The core idea here is classically taking the trace of the matrix of $L^{\dagger}\lvert \phi_{ik}(t) \rangle\langle \phi_{ik}(t)\lvert L$ involves summing over the modulus squares of all the coefficients of  the vector $L^{\dagger}\lvert \phi_{ik}(t) \rangle$, while quantum mechanically we can simply measure the probability of the final state $U_{L^{\dagger}}U_{M_k}(v_i^T,0,...,0)^T$ projecting into the first $n$-dimensional space. In other words, with the quantum method we do not have to look into individual coefficients of $L^{\dagger}\lvert \phi_{ik}(t) \rangle$, but can treat the first $n$-dimensional space of $U_{L^{\dagger}}U_{M_k}(v_i^T,0,...,0)^T$ as a whole - this is an exploitation of quantum superposition and quantum measurement. Note to obtain the probability of projecting into a subspace, a number of quantum measurements are required. This number of measurements $P$ however, does not scale with the dimension $n$ but only depends on the error that we can tolerate, as explained in Section~\ref{sec:S3}. Consequently the query complexity of the quantum method to determine $\mathrm{Tr}\big(L^{\dagger}\lvert \phi_{ik}(t) \rangle\langle \phi_{ik}(t)\lvert L\big)$ is $P$ that is a constant determined by the error tolerated, while the query complexity of any classical method cannot be reduced below $\mathcal{O}(n)$ (which is in general $\mathcal{O}(2^q)$, exponential in the number of qubits $q$) for calculating and summing over the modulus square of each individual coefficient of $L^{\dagger}\lvert \phi_{ik}(t) \rangle$. This means that the quantum method has a decisive exponential query complexity advantage over classical methods. Now if we define the total complexity of the algorithm to include the execution complexity and the query complexity, and then compare the quantum algorithm with classical methods, we see that applying $P$ measurements requires $P$ parallel implementations of the circuit for $U_{L^{\dagger}}U_{M_k}(v_i^T,0,...,0)^T$ (because multiple projection measurements cannot be applied to the same quantum state). This means that the total complexity of the quantum algorithm is the query complexity $P$ multiplied by the execution complexity $Q$, while the total complexity of a classical algorithm with the same execution complexity is $Q$ plus $\mathcal{O}(n)$. We see that if $Q$ is $\mathcal{O}(n)$ or above, the quantum algorithm does not have a total complexity advantage over classical methods. However, when $Q$ is $\mathcal{O}(\mathrm{log}(n))$ or below, which happens when the implementations of both $U_{M_k}$ and $U_{L^{\dagger}}$ are very simple, the quantum algorithm will have a decisive advantage over classical methods. An illustrative example of this scenario can be constructed in the following. Suppose the initial density matrix is given by Equation~\eqref{eq:1}  in the main text, where the physical composition of a few pure states is known. In fact, since the number of pure states in the mixture must be $\mathcal{O}(\mathrm{log}(n))$ or below for $\rho$ to be representable with a cost that scales polynomially with the qubit number, we may assume there is only one pure state $\lvert \phi \rangle = (c_1, c_2,...,c_n)$ in the density matrix for simplicity. Now also suppose there is no dynamical evolution and we are interested in the expectation value of an observable on $\phi$ itself, then the execution cost for $U_{M_k}$ is removed. Given that the system can be represented by $q$ qubits such that $2^q = n$, consider the observable: 
\begin{equation}
    \tag{S10}
    A = \sigma_z \otimes \mathbf{I}^{\otimes(q-1)} 
    \label{eq:S10}
\end{equation}
Now define $\tilde{A} = \frac{A + \mathbf{I}\lvert\lvert A \rvert\rvert}{2\lvert\lvert A \rvert\rvert}$, then the Cholesky decomposition gives:
\begin{equation}
\tag{S11}
\label{eq:S11}
L^{\dagger} = L = A = \begin{pmatrix} 1 & & & & &\\ &\ddots& & & &\\ & & 1 & & &\\ & & & 0 & &\\ & & & & \ddots & \\ & & & & & 0\end{pmatrix}
\end{equation}
and
\begin{align}
\tag{S12}
\label{eq:S12}
    \langle \tilde{A} \rangle &= \mathrm{Tr}(L^{\dagger}\lvert \phi \rangle \langle \phi \rvert L) = \sum_{i=1}^{\frac{n}{2}} \lvert c_i \rvert^2 \\
    \tag{S13}
    \label{eq:S13}
    \langle A \rangle &= 2\langle \tilde{A} \rangle -1 
\end{align}
where $c_i$’s are the coefficients of the initial state $\lvert \phi \rangle$. The sum $\sum_{i=1}^{\frac{n}{2}} \lvert c_i \rvert^2$ in Equation~\eqref{eq:S12} has a clear physical meaning of the probability of projecting into the first half space where the first qubit has the value $\lvert 0 \rangle$, and thus can be measured by a projection measurement efficiently. Note that here the observable $A$ is particularly simple that no actual implementation of $L^{\dagger}$ is required. The total complexity of the quantum algorithm evaluating $\langle A \rangle$ over $\lvert \phi \rangle$ is then only the query complexity of $P$. On the other hand, for any classical method, evaluating the sum  $\sum_{i=1}^{\frac{n}{2}} \lvert c_i \rvert^2$ costs $\mathcal{O}(n)$ steps and the total complexity is $\mathcal{O}(n) = \mathcal{O}(2^q)$. Consequently the quantum algorithm has a desicive total complexity advantage over any classical method for the example we have constructed. In addition to this special example, as discussed in the section on the generalized quantum algorithm in the main text, the Kraus operators are formulated with the basic physical processes that constitute the dynamical model, and thus typically contain only one or few non-zero elements. This means the execution cost for each $M_k$ is typically $\mathcal{O}(\mathrm{log}(n))$ or below, therefore if the number of $M_k$’s is $\mathcal{O}(\mathrm{log}(n))$ or below, and implementing the observable $A$ with $L^{\dagger}$ also costs $\mathcal{O}(\mathrm{log}(n))$ or below, we will be able to demonstrate the quantum advantage on a dynamical evolution. To find such an observable $A$ with a realistic physical meaning would be an interesting subject of a future study. 

\section{An example of the quantum circuits used to evolve the Kraus operators}
\label{sec:S5}

Below we give an example of the quantum gate sequence used to build the quantum circuit for evolving ${{M}_{1}}=\sqrt{\alpha \delta t}\left| 1 \right\rangle \left\langle  1 \right|$ at the first time step at 400 atomic unit. After being multiplied by the unitary matrix accounting for the coherent part as in Equation~\eqref{eq:17} in the main text, ${{M}_{1}}$ becomes:
\begin{equation}
\tag{S14}
\label{eq:S14}
\begin{pmatrix}
   0 & 0 & 0 & 0 & 0  \\
   0 & 0.219-0.09i & 0 & 0 & 0  \\
   0 & 0.017+0.042i & 0 & 0 & 0  \\
   0 & 0.001-0.003i & 0 & 0 & 0  \\
   0 & 0 & 0 & 0 & 0  \\
\end{pmatrix}	
\end{equation}
Using the dilation procedure described around Equation 3 in the main text this becomes a $10\times 10$ unitary matrix ${{U}_{{{M}_{1}}}}$. Using 4 qubits to cover the 10 dimensions and 2 ancilla qubits for the gate decomposition, we obtain a gate sequence of 899 gates as shown below. Note this sequence is not unique and may be further optimized, but it is good enough for the purpose of producing the results as shown in the main text when run on the IBM QASM simulator. The details of the decomposition of all the quantum circuits used are available from the corresponding author on reasonable request.

\begin{figure}[h!]
    \centering
    \includegraphics[width = 0.45\columnwidth]{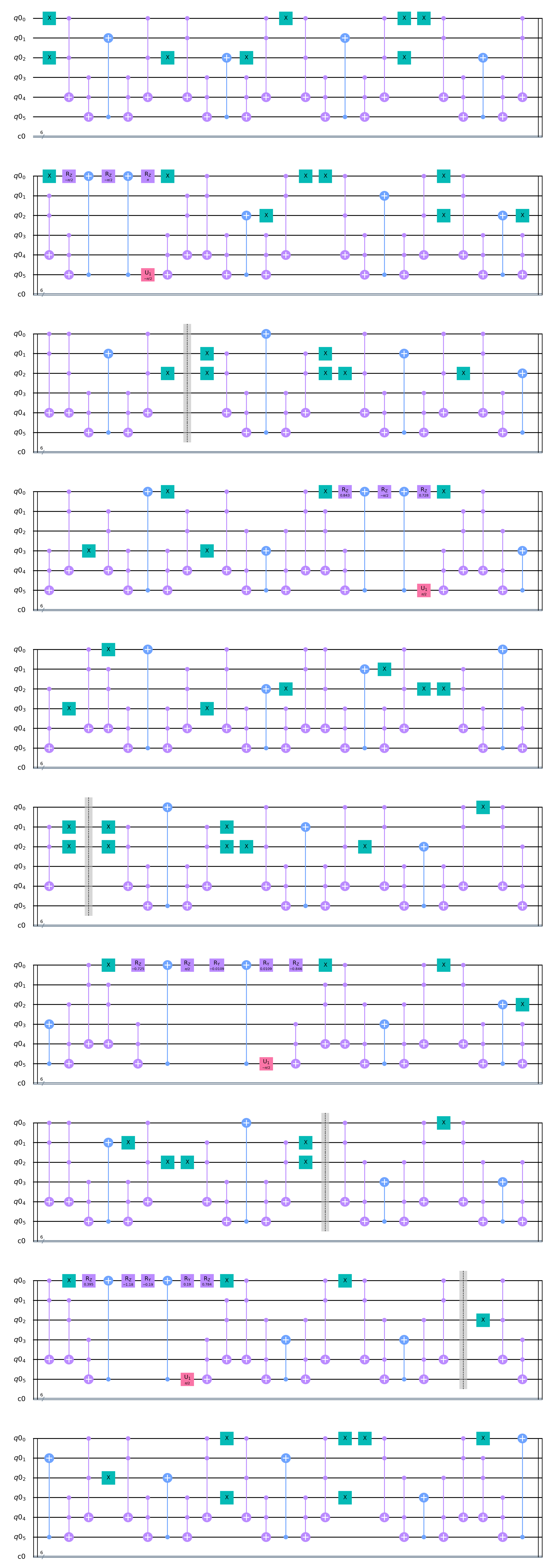}
    \includegraphics[width = 0.45\columnwidth]{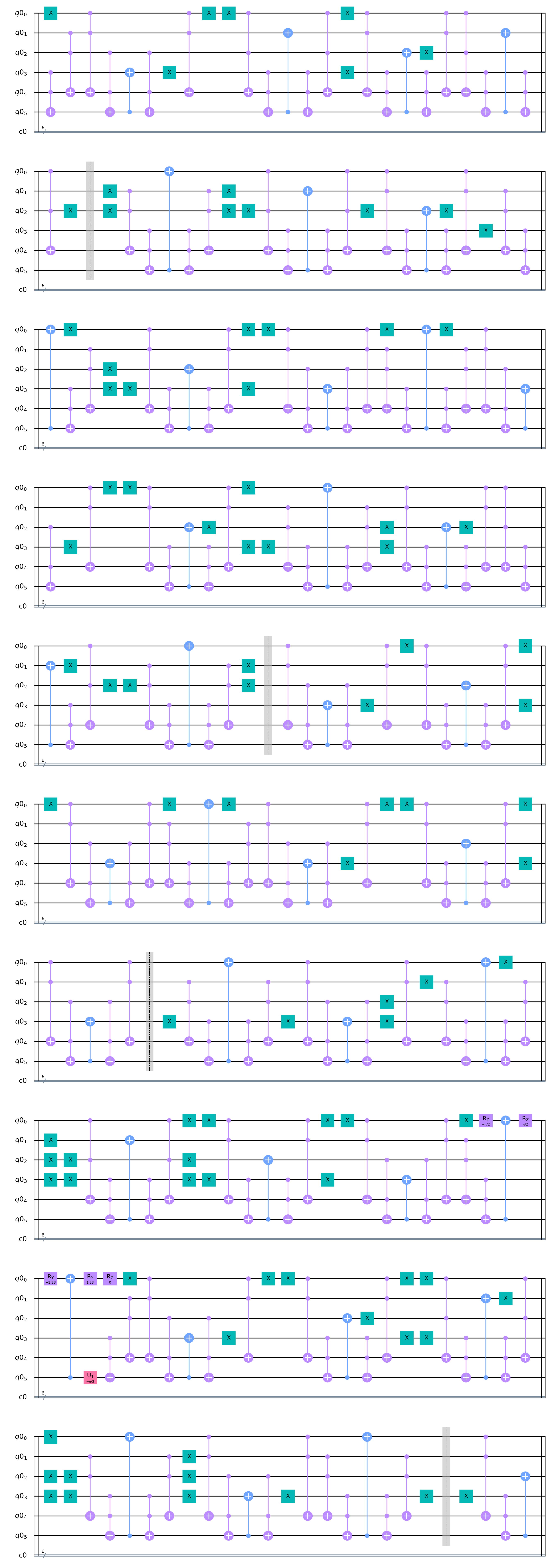}
    \label{fig:circuit1and2}
\end{figure}

\begin{figure}[h!]
    \centering
    \includegraphics[width = 0.4\columnwidth]{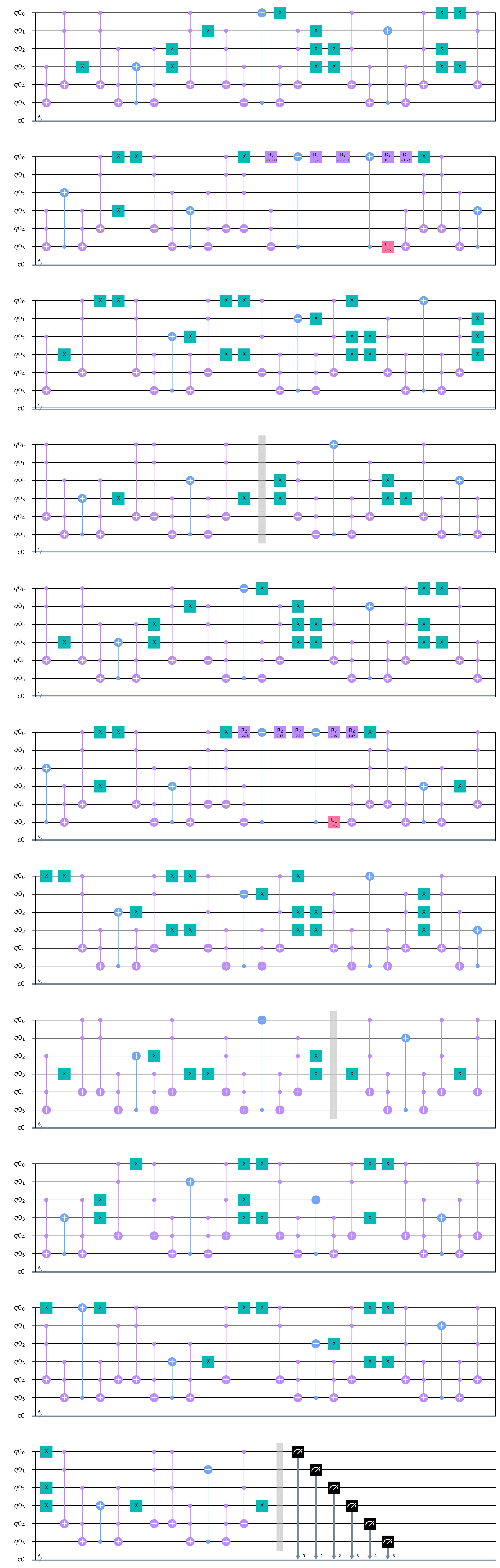}
    \caption{An example of the quantum gate sequence used to evolve a Kraus operator}
    \label{fig:circuit3}
\end{figure}


\end{document}